\documentclass[10pt]{iopart}
\usepackage[colorlinks=true,urlcolor=blue,citecolor=blue,linkcolor=red]{hyperref} 
\usepackage[margin=0.77in]{geometry}
\usepackage{graphicx}

\begin{document}
\title[]{The hidden uncertainties in particle balance calculations and their implications for assessment of plasma performance}

\author{A. A. Teplukhina$^1$, F. M. Poli$^1$, M. Podest\`a$^1$, M. Gorelenkova$^1$, G. Szepesi$^2$, Ye. O. Kazakov$^3$, Y. Baranov$^2$, \v{Z}. \v{S}tancar$^{4}$ and the JET Contributors$^*$}

\address{$^1$ Princeton Plasma Physics Laboratory, Princeton, New Jersey 08543, USA}
\address{$^2$ Culham Centre for Fusion Energy, Culham Science Centre, Abingdon, United Kingdom}
\address{$^3$ Laboratory for Plasma Physics, LPP-ERM/KMS, TEC Partner, Brussels, Belgium}
\address{$^4$ Jo\v{z}ef Stefan Institute, Ljubljana, Slovenia}
\address{$^*$ See the author list of ``Overview of JET results for optimising ITER operation'' by J. Mailloux et al. to be published in Nuclear Fusion Special issue: Overview and Summary Papers from the 28th Fusion Energy Conference (Nice, France, 10-15 May 2021)}

\ead{ateplukh@pppl.gov}
\vspace{10pt}
\begin{indented}
\item[] October 2021
\end{indented}

\begin{abstract}
Predictive simulations of mixed plasma discharges, like deuterium-tritium plasmas, rely on self-consistent models for particle transport. These models have to be validated with interpretive analysis of existing experiments. Accounting for uncertainties in the particle balance equation is important for modelling of mixed-plasma discharges because simulation results are directly affected by plasma composition. JET deuterium and mixed hydrogen-deuterium plasma discharges heated by neutral beams only are analysed with the TRANSP code. Influence of prescribed quantities and terms entering into the particle balance equation is quantitatively assessed on the computed plasma parameters, in particular on the neutron rate and the plasma energy. Large uncertainties in the prescribed ion temperature, plasma rotation and impurity content might result in a noticeable over- or underestimate of the computed neutron rate. A significant difference in the time evolution of the measured and computed neutron rate is observed in the interpretive simulations if same diffusivity for electrons and thermal ions is assumed. Increased thermal deuterium ion transport is required to reach agreement with the plasma composition measured at the plasma edge and the measured neutron rate. Thermal ion density computed with the particle balance equation is mostly affected by the ion outflux term.
\end{abstract}


\section{Introduction}
Understanding differences in particle transport properties of various plasma species is essential for analysis of present experiments and plasma scenario development. Theoretical and experimental studies on electron transport and isotope mixing have been conducted over the past years \cite{Angioni:PPCF2009, Maslov:NF2018} (and references therein) in which density peaking has been investigated in terms of plasma turbulence and balancing particle diffusion and inward convection. Isotope related turbulence suppression has been found for projections from deuterium (DD) to deuterium-tritium (DT) mixed plasma discharges \cite{Garcia:PP2018} that is important for extrapolation to ITER plasmas. Recent experimental \cite{Maslov:NF2018} and numerical \cite{Marin:NF2020} studies show that ion transport faster than electron transport can be expected in H-mode plasmas. When a self-consistent predictive model for ion transport is not available, projections to DT operation can still be done, starting from a DD discharge where the electron density profile is taken from experiments or rescaled and the plasma composition is modified to represent an equal mix of deuterium and tritium. The plasma performance calculated this way is still dependent on particle balance analysis, since the ion density profiles are computed from given electron density profile and assumptions on either the impurity content or measured plasma composition $Z_{eff}$. Uncertainties in the computed transport and in the resulting ion thermal density profiles include experimental profiles, including any missing data, and transport modeling assumptions. While electron density and temperature measurements with good time resolution are usually available in all experiments, ion temperature measurements require dedicated diagnostic neutral beams and are often acquired only over reduced time windows. Since the ion temperature is a critical measurement for the computation of the neutron yield, uncertainties may be large when the ion temperature is estimated, or taken equal to the electron temperature. This motivates the analysis described in this paper, where the effects of experimental uncertainties and user-defined transport assumptions (hidden dependencies) are discussed for two plasma discharges on JET, one with pure deuterium and one with a mix of deuterium and hydrogen. The chosen discharges are also representative of experiments on JET that miss ion temperature measurements, thus increasing the uncertainties in the computed neutron yield, which is used as a measure of the plasma performance. For analysis consistency both discharges are in L-mode with low heating by neutral beams that helps to minimise fast ion contribution to uncertainties in the calculation of the thermal ion densities.

Sensitivity studies are performed with the TRANSP code \cite{Hawryluk:PPT1980,TRANSP}, which is one of the transport solvers available for tokamak plasma discharge analysis and predictions. TRANSP features a flexible workflow and a wide choice of options for interpretive and predictive analysis of tokamak plasma discharges including advanced models for various auxiliary heating schemes and plasma transport. The TRANSP particle balance equation and its implementation to the code workflow is discussed in Section \ref{sec:TRANSP}. Section \ref{sec:DD} is dedicated to JET DD plasma modelling and sensitivity of simulation results to the TRANSP input plasma profiles. In Section \ref{sec:HD} interpretive analysis of the JET HD plasma discharge is presented along with sensitivity studies on terms entering to the TRANSP particle balance equation. Results summary and further research directions are discussed in Section \ref{sec:conclusion}.

\section{Initial assessment of terms that enter in the particle balance calculations, from experiments and from theory} \label{sec:TRANSP}

The neutron rate is an indication of plasma performance in fusion experiments; for example, for the contribution of thermonuclear reaction to the neutron rate, it depends on the deuteron density $n_{\rm{D}}$ and on the ion temperature $T_{\rm{i}}$, as $Y_{\rm{N}} = 0.5n_{\rm{D}}^2 T_{\rm{i}}^\beta$, where the parameter $\beta$ is a function of the temperature. On JET, $\beta$ was found to vary between 3.5 and 2.9 for ion temperatures between 3 and 8 keV \cite{Sasao:PPCF1994}. From a simple error propagation analysis, it is seen that a variation of 10\% on the ion temperature has a larger effect than a variation of 10\% on the density. In the absence of measurements, hypotheses are made on the magnitude of the ion temperature based on the electron temperature, which is usually available, the simplest assumption being that the two are the same. Uncertainties in the ion temperature assumptions affect the neutron rate when all nonlinearities are included. The deuteron density can be prescribed or computed from particle balance and the plasma quasi-neutrality condition, once the electron density and the plasma composition $Z_{\rm{eff}}$ are known.

Here, the free-boundary equilibrium and transport solver TRANSP \cite{Hawryluk:PPT1980,TRANSP} is used to solve the nonlinear transport equations in interpretive mode. In this operation mode, the magnetic equilibrium and the measured profiles are prescribed from the experiments and the transport equations are solved for a generic plasma observable $Q$ once the source terms $S(Q,\mathbf{r},t)$ are known, assuming that transport can be described as a diffusion/convection-like problem of the form \cite{Poli:PP2018}:
\begin{equation}
\label{eq:diffusion}
\frac{\partial Q}{\partial t} + \nabla \cdot \Gamma = S(Q,\mathbf{r},t)
\end{equation}
Here the flux $\Gamma=Q\mathbf{v}-D\nabla Q$ consists of a diffusion $D$ and a convection $\mathbf{v}$ term that are determined based on the solution of coupled equations between the ion and the electron fluids.

For each thermal ion species $s$, the global particle balance equation averaged over the plasma column is solved in TRANSP:
\begin{equation}
\frac{\partial{n_s}}{\partial{t}} = S_{bs} + G_{0s} + R_{0s} - F_s
\label{eq:particle_balance}
\end{equation}
and for the ion fluid as a whole. Here $n_s$ indicates the density of individual ion species, $S_{bs}$ represents the ion thermalisation source from neutral beam injection, $G_{0s}$ and $R_{0s}$ represent the gas and the plasma recycling sources respectively, $F_s=\nabla \cdot \Gamma_s$ implies the ion outflux and $\Gamma_s=-D\nabla n_s+n_sv_s$. Equation (\ref{eq:particle_balance}) is solved for each radial zone under the condition that quasi-neutrality is satisfied, using the plasma composition $Z_{\rm{eff}}$ as a constraint:
\begin{equation}
n_{\rm{e}} = \sum_j n_{j} Z_{j}
\label{eq:quasineutrality}
\end{equation}
\begin{equation}
n_{\rm{e}} Z_{\rm{eff}}= \sum_j n_{j} Z_{j}^2
\label{eq:zeff}
\end{equation}
where $n_{\rm{e}}$ and $Z_{\rm{eff}}$ are the electron density and the plasma effective charge, $n_{j}$, $Z_{j}$ are the individual ion species, including impurities and fast ions.

The source term $S_{bs}$ in Equation (\ref{eq:particle_balance}) is provided by the physics based code NUBEAM \cite{Pankin:CPC2004}, while the terms $G_{0s}$ and $R_{0s}$ are computed with a reduced model FRANTIC \cite{Tamor:JCP1981}. If no input is provided for the gas flow, TRANSP assumes a minimum gas flow of 1 atom per second, which is basically zero. The ion outflux $F_s$ is computed based on user-defined hypotheses on the ion transport. It will be shown in Section \ref{sec:HD} that this contribution accounts for the largest uncertainties in the computed ion densities in a multi-species plasma.

For the JET plasma discharges discussed further, the electron density profiles are prescribed from the experiments as modified hyperbolic tangent function fitting of High Resolution Thomson Scattering (HRTS) diagnostic measurements \cite{Pasqualotto:RSI2004} over the magnetic equilibrium reconstructed by the EFIT++ code \cite{Lao:NF1985,Appel:EPS2006}. If the magnetic equilibrium has been reconstructed using only magnetic measurements and is not constrained by Motional Stark Effect (MSE) measurements \cite{Hawkes:RSI1999, Stratton:RSI1999}, uncertainties in the electron density profile shape do propagate to the impurity density profiles and to the computed ion density profiles, through quasi-neutrality. The time-varying integrated plasma composition $Z_{\rm{eff}}$ is measured on JET  from visible spectroscopy (Bremsstrahlung). In the analysis described in the remainder of this paper, it is assumed that beryllium Be9 is the only impurity. The effect of uncertainties on the electron density profile and on the assumptions on $Z_{\rm{eff}}$ are discussed in Section \ref{sec:DD}.

The beam source is one of the outputs of the Monte Carlo code NUBEAM \cite{Pankin:CPC2004}. This code is used to compute neutral beam power deposition, ionization of neutral particles and thermalisation of beam ions, slowing down of fast ions and other plasma profiles related to neutral beam injection (NBI). NUBEAM takes into account beam geometry and the fractions of power corresponding to different injection energies. Beam ions contribute to the neutron rate by beam-target and beam-beam fusion reactions. The former is usually larger, therefore the computed neutron rate strongly depends on the background plasma composition.

When multiple thermal ion species are present, assumptions on the transport become the largest source of uncertainty, as it will be discussed in Section \ref{sec:HD}. In this case, it is important how the diffusivities are defined, since different ion species would have different diffusivity relative to each other. TRANSP allows to test these hypotheses by providing a number of models for the ion outflux, including computing individual diffusivities from the global ion diffusivity, fixing the species fraction, prescribing one of the species density, or selecting the values of diffusivity and conductivity. Here, a subset of models which would minimize systematic errors introduced by the end user has been selected. The first case is when both species are solved as a group and the ion transport is inferred from the electron transport. The second one is when the density of one species is prescribed and the other species is computed from particle balance and quasi-neutrality. The third case is when the density of one species is computed from diffusion and convection profiles and the density of the other species is computed from particle balance and quasi-neutrality. In all cases particle balance and quasi-neutrality are solved for and must be satisfied, but the radial flux $\Gamma_s$ in Equation (\ref{eq:particle_balance}) is computed under different assumptions. In the first case the Z-weighted summed continuity equation is solved to infer a summed flux constraint, with individual ion species evolved independently based upon their relative concentration. The diffusivity $D_s$ is an input to the model and the convective term $n_sV_s$ is computed to satisfy the global flux constraint. In the analysis discussed here, the ion diffusivity $D_0$ is assumed to be equal to the net electron diffusivity under the assumption that the electron transport is fully diffusive, $D_0=\Gamma_e/\nabla n_e$, where the electron flow velocity $\Gamma_e$ is computed from the electron balance equation averaged over the plasma column. To include isotopic effects one can set relative diffusivity of each ion species. In the case where one ion species is prescribed either through its density profile or through the transport, all known ion species are subtracted before solving for the ion outflux for the remaining species.

Since inclusion of multiple parameters with undefined uncertainties complicates the modelling task, first modelling of a single ion plasma discharge is discussed and then a mixed plasma discharge is analysed. The next section is dedicated to interpretive modelling of a deuterium JET plasma discharge in which impact of TRANSP input parameters on the computed neutron rate and plasma energy is assessed.

\section{Propagation of the uncertainties in the input parameters to the simulation output} \label{sec:DD}
The first case that is discussed is an interpretive analysis of the L-mode JET deuterium discharge \#94612. In this case, the single ion deuterium density is inferred from the electron balance equation and from the impurity density profiles, computed from $Z_{\rm{eff}}$ and from quasi-neutrality. Uncertainties in prescribed parameters, like temperature or $Z_{\rm{eff}}$ profiles, affect the simulation results through the terms entering the particle balance equation. There are two types of uncertainties. The first one is an experimental uncertainty in diagnostic measurements, and for calibrated signals it is typically within 5-10\% on JET. The second one arises when assumptions have to be made if some measurement is missing. A typical example is if the ion temperature diagnostic measurements are absent and one assumes that the ion temperature equals the electron temperature.

Uncertainty in individual parameters can have significant contribution to the simulation results. Due to nonlinear relations of different quantities in transport equations, combination of various uncertainties might compensate or enhance effect of some of them. Final contribution of all uncertainties to simulation results can be any combination due to strong non-linearity of transport equations. In this section a few quantities that contribute most to the computed neutron rate and the plasma energy, are chosen for sensitivity studies. In a typical deuterium (DD) plasma discharge with D-NBI heating, in addition to electrons and thermal D ions, plasma species include impurity ions and D beam ions. Uncertainties in the measurements and fitting procedures of such plasma parameters like the electron $T_{\rm{e}}$ and ion temperature  $T_{\rm{i}}$ profiles, electron density profiles $n_{\rm{e}}$, the plasma effective charge $Z_{\rm{eff}}$, and the plasma rotation, propagate to the computed thermal D ion density, the neutron rate and the plasma energy.

\begin{figure}[t]
\includegraphics{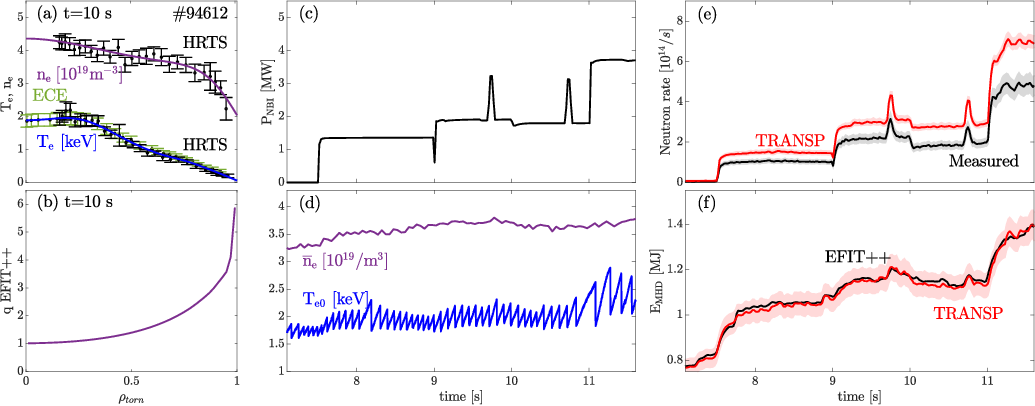}
\caption{TRANSP interpretive simulation results for {JET} \#94612 discharge: (a) fitted and measured by HRTS the electron density $n_{\rm{e}}$ at 10 s, fitted and measured by HRTS+ECE the electron temperature $T_{\rm{e}}$ profile at 10 s; (b) EFIT++ $q$ profile at 10 s; (c) the D-NBI power time trace; (d) time evolution of the line averaged electron density $\overline n_{\rm{e}}$ and central $T_{\rm{e0}}$ measured by ECE; (e) the neutron rate computed by TRANSP and measured with a 10\% technical uncertainty (grey shaded); (f) plasma MHD energy computed by TRANSP and EFIT++; the red shaded areas reflect variations in the TRANSP neutron rate and plasma energy computed with the sensitivity studies.}
\label{fig:JET_94612_overview}
\end{figure}

The discharge \#94612 features low power D-NBI heating and a long flat-top phase without significant MHD activity. This discharge has been analyzed over the time window from 7 s to 11.6 s, during which the neutron beam power is increased from 0 to 3.8 MW. The plasma current $I_{\rm{p}}=2.5$ MA and the toroidal magnetic field $B_0=2.9$ T are kept constant during the considered time period. Electron temperature $T_{\rm{e}}$ profiles are known from electron cyclotron emission (ECE) \cite{delaLuna:RSI2004} and HRTS measurements. Electron density $n_{\rm{e}}$ profiles are measured by HRTS. Time evolution of plasma equilibrium and $q$ profiles are provided by the kinetic equilibrium reconstruction code EFIT++. To improve consistency in computed plasma equilibrium and plasma profiles, three iterations have been performed in the TRANSP-EFIT++ loop. TRANSP plasma profiles are used by EFIT++ to reconstruct plasma equilibrium and profiles are remapped to the updated equilibrium after each iteration. Note that $q$ profiles are computed by EFIT++ too but they are not constrained by MSE diagnostic measurements. Figure \ref{fig:JET_94612_overview}(a)--(d) shows the profiles of electron density $n_{\rm{e}}$ and temperature $T_{\rm{e}}$, safety factor $q$ at 10 s, the time traces of the NBI input power, ECE $T_{\rm{e}}$ at the plasma axis, the line averaged electron density $\overline n_{\rm{e}}$. Density profiles are fitted by a modified hyperbolic tangent function fit and electron temperature profiles are fitted by a cubic spline fit. Due to the lack of impurity diagnostic measurements, beryllium Be9 is determined as the single plasma impurity. The plasma effective charge $Z_{\rm{eff}}$ profiles are assumed to be flat and fixed at $1.2$ which is in close agreement with the experimental time trace. Since ion temperature $T_{\rm{i}}$ diagnostic measurements are not available for \#94612, we assume $T_{\rm{i}}=T_{\rm{e}}$.

The simulation results of the interpretive TRANSP run are shown in Figure \ref{fig:JET_94612_overview}(e)--(f). The computed neutron rate exceeds the measured rate by 40\%, whereas the TRANSP plasma energy is in good agreement with the EFIT++ data. In the discussed case, beam ions slow down quickly (less than 70 ms for full energy D-NBI ions) and more than 90\% of neutrons are produced by the beam-target fusion reactions. For modelling JET L-mode plasma discharges with low NBI power, inconsistencies between the neutron rate computed by TRANSP and measured one have been reported in \cite{Weisen:NF2017}, with a difference of 50\%-100\% between the measured and computed neutron rate in the JET baseline and hybrid discharges. Authors focused on NBI modelling properties trying to explain the difference, however the source of inconsistency was not found. Here, we go beyond neutral beam calculations and assess a broader range of parameters that can affect the computed neutron rate and that are directly related to the particle balance equation.

A good agreement in the computed and experimental plasma energy but not in the neutron rate might indicate large uncertainties in $T_{\rm{i}}$, $Z_{\rm{eff}}$ or plasma rotation profiles. Profile diagnostic measurements are not available for these quantities, therefore $T_{\rm{i}}$=$T_{\rm{e}}$, flat $Z_{\rm{eff}}$ profiles at 1.2 and zero plasma rotation have been used in the simulation. Propagation of uncertainties in the prescribed profiles to the computed parameters is assessed in terms of the D line averaged density $\overline n_{\rm{D}}$, the neutron rate and the plasma energy. The results of sensitivity studies are summarised in Table \ref{table:JET_94612_sens}.
\begin{figure}[t]
\includegraphics{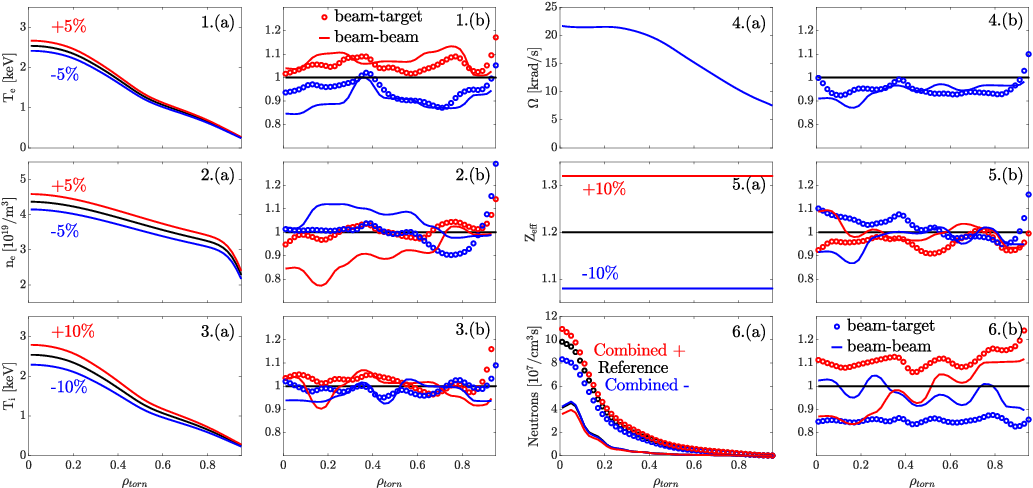}
\caption{Sensitivity studies of TRANSP computed beam-target and beam-beam neutron rate densities. Subfigures (a) present the test plasma profiles, subfigures (b) present the ratio of the beam-target (circles) and beam-beam (solid) neutron rate density computed for the reference and test cases. 1(a)--(b) 5\% variation in the electron temperature $T_{\rm{e}}$; 2(a)--(b)  5\% variation in the electron density $n_{\rm{e}}$; 3(a)--(b) 10\% variation in the ion temperature $T_{\rm{i}}$; 4(a)--(b) non-zero plasma rotation $\Omega_0=20$ [krad/s] and $\Omega_{\rm{edge}}=5$ [krad/s]; 5(a)--(b) 10\% variation in the ion temperature $Z_{\rm{eff}}$; 6.(a)--(b) combined uncertainties $T_{\rm{e}} \pm 5\%$, $n_{\rm{e}} \pm 5\%$, $T_{\rm{i}} \pm 10\%$, $Z_{\rm{eff}}\mp 10\%$, $\Omega_0=20$ [krad/s] and $\Omega_{\rm{edge}}=5$. Plasma profiles are sampled at 11.64 s.}
\label{fig:JET_94612_profiels}
\end{figure}
\begin{table}
\begin{tabular}{ |c|c|c|c| } 
\hline
  & $\overline n_{\rm{D}}$, \% & Neutron rate, \% & $E_{\rm{MHD}}$, \% \\
\hline
$T_{\rm{e}} \pm 5\%$ & -- & $\pm 5$ & $\pm 3$ \\
\hline
$n_{\rm{e}} \pm 5\%$ & $\pm$ 5 & $<1$ & $\pm 5$ \\
\hline
$T_{\rm{i}} \pm 10\%$ & -- & $\pm 6$ & $\pm 5$ \\
\hline
$Z_{\rm{eff}} \mp 10\%$ & $\pm 4$ & $\pm 4$ & $<2$ \\
\hline
$\Omega_0=20$ krad/s & -- & -5 & -- \\
\hline
$Combined$ & $\pm 9$ & -18, +16 & -12, +13 \\
\hline
\end{tabular}
\caption{Summary on the TRANSP sensitivity studies for {JET} \#94612 interpretive analysis.}
\label{table:JET_94612_sens}
\end{table}

ECE and HRTS diagnostic error bars are within 5\% for $T_{\rm{e}}$ and $n_{\rm{e}}$. A 5\% scaling of the prescribed $T_{\rm{e}}$ profiles, as shown in Figure \ref{fig:JET_94612_profiels}.1(a), results in 8\% and 5\% variation in the computed neutron rate and the plasma energy $E_{\rm{MHD}}$. Neutrons are produced by DD beam-target and beam-beam fusion reactions, and electron temperature affects both fusion rates. The beam-target $R_{\rm{bt}}$ and beam-beam $R_{\rm{bb}}$ neutron rate profiles, sampled at 11.64 s for $T_{\rm{e}}$ 5\% scaling tests, are normalised to the reference values and shown in Figure \ref{fig:JET_94612_profiels}.1(b). As expected, $R_{\rm{bt}}$ and $R_{\rm{bb}}$ increase with $T_{\rm{e}}$ and vice versa. Similar scaling of the electron density profiles affects the thermal ion density and the plasma energy. In more dense plasmas beam ions are slowing down faster, thus $R_{\rm{bb}}$ is decreased (Figure \ref{fig:JET_94612_profiels}.2(b)). Note that the plasma energy is more affected by the electron density, whereas the neutron rate is sensitive to the electron temperature.

An assumption on similar electron and ion temperatures may result in an underestimate of the neutron rate, since the beam-target and beam-beam cross section increases with $T_{\rm{i}}$ as shown in Figure \ref{fig:JET_94612_profiels}.3(a)-(b). CXRS error bars of 10\% are typically observed for JET plasmas, therefore 10\% variation in $T_{\rm{i}}$ is used in the sensitivity tests. Uncertainties in $T_{\rm{i}}$ contribute more to the neutron rate than to the plasma energy (Table \ref{table:JET_94612_sens}). In the considered discharge NBI power is absorbed mostly by electrons. Therefore the initial assumption $T_{\rm{i}}$=$T_{\rm{e}}$ might not be correct at certain phases of the plasma discharge, and one can expect $T_{\rm{i}}$ to be lower than $T_{\rm{e}}$ during the NBI heating phase. It is difficult to estimate differences in ion and electron temperatures without $T_{\rm{i}}$ diagnostic measurements. However, it should be taken into account that an assumption $T_{\rm{i}}$=$T_{\rm{e}}$ in case of stronger electron heating might introduce uncertainties larger than 10\%.

Profiles for plasma rotation associated with neutral beam heating are not available for this discharge, therefore plasma rotation is not included to the reference TRANSP run. To estimate its influence on the neutron yield, an artificial plasma angular rotation has been introduced with $\Omega_{\rm{axis}}=20$ krad/s and $\Omega_{\rm{edge}}=5.0$ krad/s which velocity profile is shown in Figure \ref{fig:JET_94612_profiels}.4(a). The plasma rotation on the axis is determined according to JET data reported in \cite{Nave:EPS2007}. It is used only as a test case, since the diagnostic data have been obtained for experiments with the JET carbon wall. The rotation profile is fixed in time, except during the ohmic phase where the plasma rotation is assumed to be zero. Influence of such slow rotation on the neutron rate is comparable with corresponding $T_{\rm{e}}$ and $n_{\rm{e}}$ contribution. If the plasma rotation is included, the beam ion relative velocity is reduced and NBI deposition profiles are broadened, therefore the beam-beam and beam-target neutron rates are decreased (Figure \ref{fig:JET_94612_profiels}.4(b)).

In the absence of profile measurements, the most simple way to include impurities is to use measured integrated $Z_{\rm{eff}}$ assuming flat profiles. Experimental uncertainty of 10\% in $Z_{\rm{eff}}$ leads to 5\% and $<2\%$ variation in the neutron rate and plasma energy respectively through modified distribution of thermal D ions in the plasma volume. One can see from profiles shown in Figure \ref{fig:JET_94612_profiels}.5(a)--(b) that higher $Z_{\rm{eff}}$ results in lower $R_{\rm{bt}}$ due to lower D thermal ion density and a fewer number of beam-target fusion reactions. Larger effect on the computed quantities might be expected if non-flat $Z_{\rm{eff}}$ profiles are included due changes in local particle transport.

Even though uncertainties in each parameter have a small impact on the computed neutron rate, that is within 10\% technical uncertainty for the measured neutrons, a combination of such uncertainties can lead to more noticeable changes. A TRANSP run with the input profiles modified like $T_{\rm{e}} \pm 5\%$, $n_{\rm{e}} \pm 5\%$, $\Omega_0=20$ [krad/s] and $\Omega_{\rm{edge}}=5$, $T_{\rm{i}} \pm 10\%$, $Z_{\rm{eff}} \mp 10\%$ results in a 18\% drop or a 16\% increase in the computed neutron rate (Table \ref{table:JET_94612_sens}). The beam-beam and beam-target neutron rate density profiles, shown in \ref{fig:JET_94612_profiels}.6(a)--(b), are noticeably affected in the entire plasma volume. It has been found that a 5--10\% variation in the TRANSP input parameters does not explain the 40\% overestimate in the neutron rate. However, if uncertainties in multiple input parameters are included, it can cause a deviation in the computed neutron rate for more than 10\%. Prescribed electron temperature and density affect both the computed neutron rate and the plasma energy. However, the plasma energy is less sensitive to $T_{\rm{i}}$, $Z_{\rm{eff}}$ and plasma rotation. Larger uncertainties in these parameters than in $T_{\rm{e}}$ and $n_{\rm{e}}$ can be expected due to absence of corresponding profile diagnostic measurements.

\section{Modelling of a plasma discharge with multiple ion species} \label{sec:HD}
In mixed plasma discharges with prescribed profiles based on diagnostic measurements of high quality, one still can expect some uncertainties coming from a thermal ion transport model. If electron density profiles are known from the experiment then the electron diffusivity can be computed with the particle balance equation. Since density measurements for ions are typically not available, the most natural assumption is to approximate that the ion diffusivity is equal to the electron diffusivity. The ion densities are then computed from particle balance assuming quasi-neutrality. However, this assumption might fail in the presence of multiple ion species, as discussed in this section. The thermal ion particle balance equation is solved for the total thermal ion density to ensure the quasi-neutrality condition and $Z_{\rm{eff}}$. To compute individual thermal ion densities the particle balance equation is solved for each of them. To split the total thermal ion density between different species one needs to specify transport coefficients, as well as gas and recycling sources. In this section it is discussed how the choice of a thermal ion transport model and plasma recycling affects computed ion densities and the neutron rate.

\begin{figure}[t]
\includegraphics{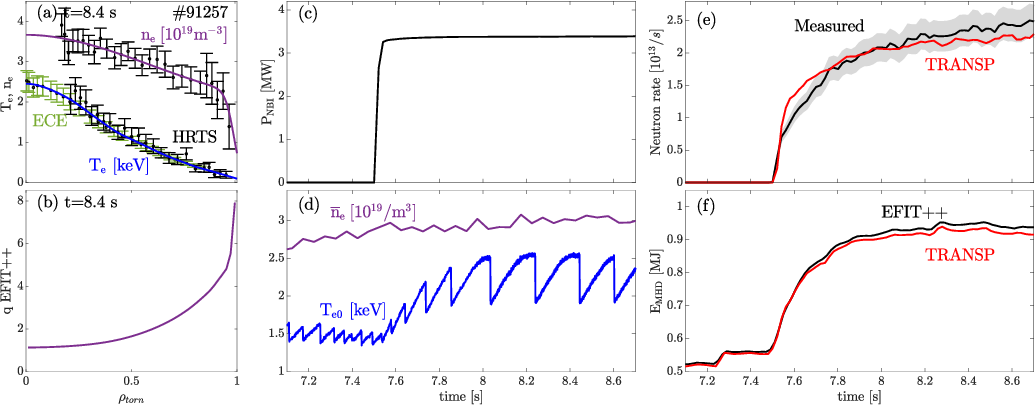}
\caption{TRANSP interpretive simulation results for {JET} \#91257 discharge: (a) fitted and measured by HRTS the electron density $n_{\rm{e}}$ at 8.4 s, fitted and measured by HRTS+ECE the electron temperature $T_{\rm{e}}$ profile at 8.4 s; (b) EFIT++ $q$ profile at 8.4 s; (c) the D-NBI power time trace; (d) time evolution of the line averaged electron density $\overline n_{\rm{e}}$ and central $T_{\rm{e0}}$ measured by ECE; (e) the neutron rate computed by TRANSP and measured with a 10\% technical uncertainty (grey shaded); (f) plasma MHD energy computed by TRANSP and EFIT++.}
\label{fig:JET_91257_overview}
\end{figure}
For interpretive analysis with TRANSP we have chosen a JET mixed deuterium and hydrogen (HD) plasma discharge \#91257 with D-NBI heating. Overview of the discharge for the time window 7.1 s--8.5 s is presented in Figure \ref{fig:JET_91257_overview}. Fitted HRTS and ECE diagnostic measurements are used to prescribe the electron temperature $T_{\rm{e}}$ and density (HRTS only) profiles $n_{\rm{e}}$. Similar to \#94612, plasma equilibrium data, including $q$ profile time evolution, are computed by EFIT++ (without the MSE constraint) and are consistent with the TRANSP pressure profiles. On-axis D-NBI injectors operate at 3.4 MW in total starting from 7.5 s. The plasma current and the toroidal magnetic field are kept constant at 2.0 MA and 2.85 T correspondingly. Same as for \#94612, we assume $T_{\rm{i}}=T_{\rm{e}}$ and $Z_{\rm{eff}}=1.2$ with Be9 as plasma impurity. There is a low variation in the line averaged electron density $\overline n_{\rm{e}}$. Regular sawtooth crashes are observed starting from 7.8 s as captured by the central ECE signal in Figure \ref{fig:JET_91257_overview}(d). To focus on assessment of uncertainties in the computed plasma composition, modelling of sawtooth crashes is not included to the present analysis.

With prescribed $n_{\rm{e}}$ and $Z_{\rm{eff}}$ one can compute densities of impurity $n_{\rm{Be}}$ and thermal ions $n_{\rm{i}}$ using Equations (\ref{eq:particle_balance})--(\ref{eq:zeff}). However, this input information is not sufficient to determine densities of individual thermal ion species, H and D. One of the solutions is to prescribe density of one of the ion species as a fraction of $n_{\rm{e}}$. Usually for JET plasmas, a relative ratio of H $n_{\rm{H}}$ and D $n_{\rm{D}}$ ion densities is known from the spectroscopic measurements at the plasma edge. Ion density $n_{\rm{D}}$ is prescribed as $n_{\rm{D}} = (0.03 \div 0.047) \cdot n_{\rm{e}}$ in order to satisfy a 3.2--5\% increase in measured $n_{\rm{D}}/(n_{\rm{D}}+n_{\rm{H}})$. The spectroscopic measurements of the H/D ratio with uncertainty of 11\% are taken at the outer divertor area. This assumption is based on the measurement at the edge and fraction is taken across the entire profile. In TRANSP simulation with such an additional constraint on the thermal ion density good agreement is observed for the neutron rate computed by TRANSP and experimentally measured, as well as for the plasma energy computed by TRANSP and EFIT++, as shown in Figure \ref{fig:JET_91257_overview}(e)--(f). For the neutron rate much better agreement has been achieved than for the DD discharge in Section \ref{sec:DD}. Since the analysed discharges have been performed in different years (2016 for \#91257 and 2019 for \#94612), it is complicated to estimate differences in the machine state, diagnostic calibration, and their contribution to the computed quantities. Due to the high nonlinearity of the transport models, uncertainties coming from different sources can compensate for each other. For the presented analysis of \#91257, in particular, $n_{\rm{D}}$ is prescribed from the signal taken at the divertor region. Therefore, firstly, its time evolution is not computed consistently with the NBI thermalisation source and, secondly, the H/D ratio in the core region might be different from the divertor one. As discussed earlier in Section \ref{sec:DD}, uncertainties in $T_{\rm{i}}$, $Z_{\rm{eff}}$ and neglecting plasma rotation contribute to the computed neutron rate too.

In the absence of isotopic ratio measurements, for predictive studies for example, one can specify only initial thermal ion density profiles and to solve the particle balance equation to compute their time evolution. In this case, a reliable transport model is required. Note that the plasma energy does not depend on the density of individual thermal ion species but on the total thermal ion density that is determined by the electron density and $Z_{\rm{eff}}$. Both parameters are prescribed and therefore are not modified in the simulations discussed below.

TRANSP initial densities of thermal ions are fixed at 96\% of H and 4\% of D. Figure \ref{fig:JET_91257_diff_recyc_tests} shows the computed time traces of the relative concentration of D ions averaged in the core $R_{\rm{D}} = \overline n_{\rm{D}}/(\overline n_{\rm{D}}+ \overline n_{\rm{H}})$, the neutron rate and the thermal D diffusivity profile at 8.6 s. The parameter $R_{\rm{D}}$ reaches 20\% at 8.6 s versus 5\% measured at the plasma edge. Since most of neutrons are produced by the DD beam-target fusion reaction, the computed neutron rate is increased more than four times compared to the simulation with prescribed $n_{\rm{D}}$. As shown in Figure \ref{fig:JET_91257_diff_recyc_tests}(c) for the time slice 8.6 s, thermal D ion diffusivity computed with the prescribed $n_{\rm{D}}$ is significantly larger than $D_{\rm{D}}=D_{\rm{e}}$. The observed difference between the measured neutron rate, the HD density ratio, and the same parameters computed by TRANSP indicates that the assumption on similarity of thermal ion and electron transport is not valid for the discussed case. One of possible explanations is that there is a strong central source of thermalised D-NBI ions. NUBEAM considers beam ions as thermalised when their temperature reaches $3/2 \cdot \rm{max}(T_{\rm{i}},\; T_{\rm{e}})$, i.e. D ion energy, and consequently their diffusivity, is higher than for background ions and electrons. In \#91257 the initial content of thermal D ions is low and comparable with D-NBI thermalisation source, therefore their transport can not be completely characterised by transport of background species like electrons. In case of plasma discharges with a more balanced plasma composition, like 50\%-50\% DT plasma discharges, contribution of thermalised ions might be less noticeable.

Further understanding of ion transport models will require spectroscopic measurements of the plasma composition in the plasma core. One may refer to gyrokinetic simulations to get an estimate of thermal ion diffusivity. With present sensitivity studies, one can conclude that thermal ion transport increased in comparison to transport of electrons can be expected in plasmas with a strong core source of thermal ions. It is important to acknowledge that using additional constraints on thermal ion densities in mixed plasma discharges, like prescribed as a fraction of $n_{\rm{e}}$, does not allow to reproduce their time evolution consistently with beam ion thermalisation, ion outflux, gas and recycling sources. Sensitivity of the computed thermal ion density on these terms is discussed below.

\subsection{Sensitivity of individual terms in the particle balance equation} \label{subsec:sens}
There are three terms that contribute to particle balance: beam ion thermalisation source, the wall source and the ion outflux. The first one is determined by a certain NBI heating scenario, in particular by beam geometry and deposition. These quantities are machine depended, they are used as direct inputs to NUBEAM and are not controlled by a user during numerical analysis. Other uncertainties in the thermalisation source are small and related to losses at the SOL and the plasma edge, such as charge exchange, shine-through and fast ion orbit losses. In the discussed case total losses of 4\% are estimated that includes 2\% of shine-through loss. Larger uncertainties are expected in two other terms, the ion outflux and the wall source. The former is determined according to a transport model chosen by a user. The latter represents a particle source inside the SOL, however it is computed from the diagnostic signal at the divertor area and is combined with a neutral gas source.

\begin{figure}[t]
\includegraphics{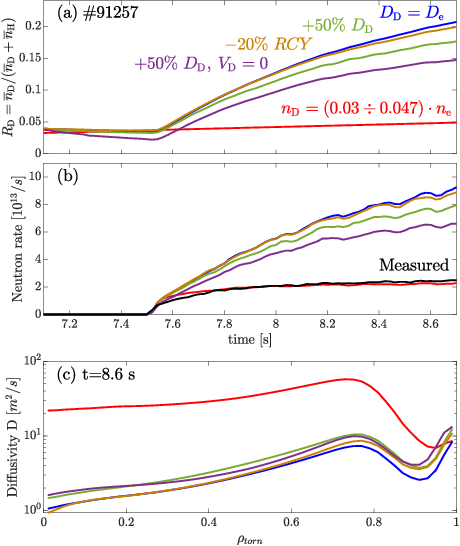}
\caption{Sensitivity runs for \#91257 with different models used to compute D thermal ion density $n_{\rm{D}}$. (a) the time trace of the line averaged density ratio $R_{\rm{D}}=\overline n_{\rm{D}}/(\overline n_{\rm{D}}+\overline n_{\rm{H}})$; (b) the measured neutron rate and computed by TRANSP; (c) thermal D ion diffusivity $D_{\rm{D}}$ sampled at 8.6 s.}
\label{fig:JET_91257_diff_recyc_tests}
\end{figure}
\begin{figure}[t]
\includegraphics{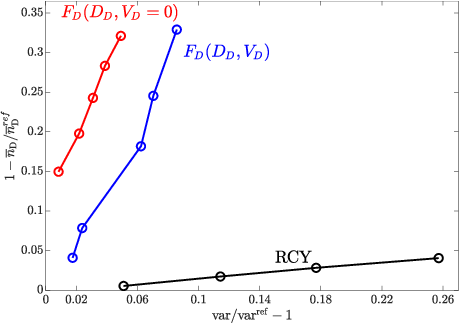}
\caption{Variation in the ion outflux $F_{\rm{D}}$ and recycling $RCY$ terms versus the line averaged density $\overline n_{\rm{D}}$. Quantities $var^{ref}$ and $var$ correspond to the ion outflux or the recycling terms computed for the reference and test cases. The ion outflux is computed for the diffusive-convective $F_{\rm{D}}(D_{\rm{D}},V_{\rm{D}})$ and diffusive only $F_{\rm{D}}(D_{\rm{D}},V_{\rm{D}}=0)$ cases. All quantities are sampled at 8.6 s.}
\label{fig:JET_91257_sensitivity_summary}
\end{figure}

The ion outflux term in the particle balance equation is defined as a combination of diffusive and convective fluxes. An increase of $D_{\rm{D}}$ by 50\% compared to the simulation with $D_{\rm{D}}=D_{\rm{e}}$ results in a 3\% and 10\% decrease in $R_{\rm{D}}$ and the neutron rate correspondingly. Corresponding time traces and diffusivity profiles are shown in Figure \ref{fig:JET_91257_diff_recyc_tests}. If one assumes only diffusive transport, i.e. convectivity of thermal D ions $V_{\rm{D}}=0$, $R_{\rm{D}}$ and the neutron rate are decreased for 5\% and 20\%.

In \#91257 the input gas is only $\rm{H}_2$, therefore the D wall source is modified through the plasma recycling term. The $\rm{D}_{\alpha}$ diagnostic signal is used to compute the recycling source in simulations. For JET plasmas $\rm{D}_{\alpha}$ measurements are taken by the visible spectroscopy diagnostic at the divertor. Since the SOL is not included to particle balance, the recycling source brings additional uncertainty to particle balance, thus to the simulation results. The plasma recycling affects not only the edge particle flux but global particle balance, thus the plasma composition and the neutron rate. However, overall effect is not significant, and a 20\% decrease in recycling results in a less than 5\% decrease in $R_{\rm{D}}$ and the neutron rate (Figure \ref{fig:JET_91257_diff_recyc_tests}).

Sensitivity studies have been performed to asses quantitatively the influence of the ion outflux and the wall source, here the recycling term only, on the computed line-averaged D thermal ion density $\overline n_{\rm{D}}$. The ion outflux term is computed in two forms: 1) a sum of diffusive and convective terms and 2) a diffusive term only, i.e $V_{\rm{D}}=0$. Results are reported in Figure \ref{fig:JET_91257_sensitivity_summary} where $var^{ref}$ and $var$ refer to the ion outflux or the recycling terms computed at 8.6 s for the reference and test cases. TRANSP interpretative run with $D_{\rm{D}} = D_{\rm{e}}$ is used as a reference for the discussed sensitivity runs. In case of the diffusive-convective ion outflux, $F_{\rm{D}}(D_{\rm{D}},V_{\rm{D}})$, $D_{\rm{D}}$ is scaled as $(1.1 \div 2.0) \cdot D_{\rm{D}}^{ref}$. As shown in Figure \ref{fig:JET_91257_sensitivity_summary}, an 8\% increase in the ion outflux results in a more than 30\% decrease in $\overline n_{\rm{D}}$, consequently in the computed neutron rate. The convective term is computed by TRANSP in order to satisfy the local quasi-neutrality and $Z_{\rm{eff}}$ conditions. That is a nonlinear loop in which both H and D densities are involved and it partially masks contribution of the diffusive term to the ion outflux. If the convective term is excluded, $F_{\rm{D}}(D_{\rm{D}},V_{\rm{D}}=0)$, then only a 3\% increase in the ion outflux term is needed to reach a 30\% decrease in $\overline n_{\rm{D}}$. Impact of plasma recycling on $\overline n_{\rm{D}}$ and the computed neutron rate is much less significant. For the sensitivity studies plasma recycling is varied between 95\%-75\% of the reference value. A 25\% decrease in the recycling terms results in only a 5\% drop of $\overline n_{\rm{D}}$. The ion outflux term compared to the other terms in the particle balance equation has the strongest impact on the computed thermal ion density.

\section{Conclusion} \label{sec:conclusion}
Projections from DD to DT plasma discharges require reliable predictions of the plasma composition and its time evolution, thus sophisticated and self-consistent particle transport models. In the absence of such models, extrapolations rely on particle balance calculations for a given electron density profiles -- which might be rescaled from deuterium plasmas -- under given hypotheses for the ion diffusivity. In the presence of multi-species plasmas, this can be a large source of uncertainties. In order to assess the effect of individual terms in the particle balance calculations on the computed neutron rate, the TRANSP solver has been used here to test various hypotheses in a self-consistent framework.

It has been discussed in Section \ref{sec:DD} that validated diagnostic data are essential to get reliable simulation results, and assessment of uncertainties coming from diagnostic measurements should be done. Variation of 5-10\% in the input plasma profiles, like the electron temperature or the plasma effective charge, can lead to a similar discrepancy in the computed neutron rate and the plasma energy, as shown in Figure \ref{fig:JET_94612_overview}. Lack of validated diagnostic data might bring non-negligible uncertainty in the simulation results and should be taken into account for experimental planning or evaluation of plasma performance.

With profiles prescribed from high quality measurements there are uncertainties coming from the ion transport models. Decoupling measurements and model contribution to the simulation results is a challenging task due to strong non-linearity of transport equations. Particle balance is sensitive to the ion outflux, beam ion thermalisation, recycling and input gas terms. It has been discussed in Section \ref{sec:HD} that an assumption on similar ion and electron diffusivity might not be correct in certain cases. For the JET plasma discharge \#91257 analyzed here, increased transport of thermal D ions is expected according to sensitivity studies on the computed plasma composition and comparison of the computed and measured neutron rates. Among other terms in the particle balance equation, the ion outflux has the largest effect on the computed ion densities when multiple background species are present.

Interpretive transport analysis and sensitivity studies presented in Sections \ref{sec:DD}--\ref{sec:HD} indentify the range of uncertainties that can be expected in the neutron rate and plasma energy depending on the transport settings and input plasma profiles. There are several other aspects important for reliable interpretive analysis that have not been investigated in details but still might have non-negligible impact on the plasma parameters. One of them is quality of plasma equilibrium reconstruction that depends on availability of internal measurements of current profile. In the discussed analysis of JET \#91257 and \#94612 plasma equilibrium, including time-dependent q profiles, was prescribed by EFIT++ not constrained by the MSE diagnostic. EFIT++ reconstructs magnetic poloidal flux and toroidal current density profiles based on external and internal magnetic measurements. In the discussed cases there are no additional constraints on the internal current density profiles coming from local measurements of pitch angels of the magnetic field lines by the MSE diagnostic. Thus, some uncertainty propagates to the simulation results through the prescribed q profiles and electron temperature and density profiles mapped on the selected plasma equilibrium. Time evolution of thermal D ion density computed by TRANSP with the particle balance equation takes into account the NBI particle source and charge exchange losses provided by NUBEAM. Sawtooth crashes are not included to the the JET \#91257 interpretive analysis discussed in Section \ref{sec:HD}. However, they can affect the neutron yield through modifications in the fast ion distribution. Such an MHD activity is localized in the plasma core for a short period of time and could modify the neutron yield by a few percent.

\section*{Acknowledgments} \label{sec:ackn}
The authors thank J. Garcia, K. Kirov and M. Maslov for valuable discussions.

This work was supported by the U.S. Department of Energy under contract number DE-AC02-09CH11466. The United States Government retains a non-exclusive, paid-up, irrevocable, world-wide license to publish or reproduce the published form of this manuscript, or allow others to do so, for United States Government purposes. This work has been carried out within the framework of the EUROfusion Consortium and has received funding from the Euratom research and training programme 2014-2018 and 2019-2020 under grant agreement No 633053. The views and opinions expressed herein do not necessarily reflect those of the European Commission. The data that support the findings of this study are available from the corresponding author upon reasonable request.

\section*{References}
\bibliographystyle{ieeetr}
\bibliography{bibliography}

\end{document}